\documentclass[notoc]{JHEP3}
\usepackage{amssymb,amsmath}
\usepackage{latexsym}

\def\eps{\epsilon}

\def\d{\partial}
\def\cA{{\cal A}}

\def\cF{{\cal F}}

\def\cI{{\cal I}}

\def\cK{{\cal K}}
\def\cL{{\cal L}}

\def\cN{{\cal N}}

\def\cQ{{\cal Q}}

\def \eps{{\epsilon}}

\newcommand{\abs}[1]{\lvert#1\rvert}

\def\buildrel#1_#2^#3{\mathrel{\mathop{\kern 0pt#1}\limits_{#2}^{#3}}}

\newcommand{\SL}{\mbox{SL}(2,\mathbb{R})}

\def \dv{{\text{d}_V}}

\title{Centrally extended symmetry algebra of asymptotically G\"odel spacetimes}

\author{Geoffrey Comp\`ere \\ {Physique Th\'eorique et Math\'ematique,
  Universit\'e Libre de Bruxelles and International Solvay
  Institutes, Campus Plaine C.P. 231, B-1050 Bruxelles, Belgium \\ E-mail: \email{gcompere@ulb.ac.be}}}

\author{St\'ephane Detournay \\ Dipartimento di Fisica, Universit\`a degli Studi di Milano,
Via Celoria 16, 20133 Milano, Italy \\ Email:
\email{stephane.detournay@mi.infn.it}}

\dedicated{Dedicated to Noah and Chantal}

\date{\today}

\abstract{We define an asymptotic symmetry algebra for
three-dimensional G\"odel spacetimes supported by a gauge field
which turns out to be the semi-direct sum of the diffeomorphisms
on the circle with two loop algebras. A class of fields admitting
this asymptotic symmetry algebra and leading to well-defined
conserved charges is found. The covariant Poisson bracket of the
conserved charges is then shown to be centrally extended to the
semi-direct sum of a Virasoro algebra and two affine algebras.

\bigskip\bigskip}

\preprint{ULB-TH/07-01, IFUM-885-FT\ hep-th/0701039}


\begin{document}

\section{Motivations}

General relativity, although providing an elegant and very
satisfactory classical description of the gravitational
interaction, has left us with numbers of conceptual physical
issues. One of them is related to the existence of black hole
solutions, and the fact that these seem to be naturally endowed,
through the laws of black hole mechanics, with a macroscopic
entropy equal to the quarter of the area of their event horizon.
It has therefore since then been a challenge for candidates to a
quantum theory of gravity to reproduce, from first-principles,
this entropy by a counting of micro-states. By now, a large number
of such derivations have appeared within many distinct frameworks,
e.g. in string theory or loop quantum gravity (see e.g.
\cite{Rovelli:1997qj}).

Another puzzle relies on the observation that some geometries
arising from Einstein's general relativity exhibit closed
time-like curves. These include namely the G\"odel universe
\cite{Godel:1949ga}, the Gott time-machine \cite{Gott:1990zr} and
the region behind the inner horizon of Kerr black holes. Since the
presence of closed time-like curves signals a strong breakdown of
causality, Hawking advocated through his chronology protection
conjecture that ultraviolet processes should prevent such
geometries from forming \cite{Hawking:1991nk}. The implications of
this proposition have namely been addressed in context of string
theory in a series of works (see e.g.
\cite{Dyson:2003zn,Israel:2003cx,Johnson:2004zq}, and also
\cite{Costa:2005ej} for an extensive list of references). Also,
higher-dimensional highly supersymmetric G\"odel-like solutions
were found in supergravity \cite{Gauntlett:2002nw,Harmark:2003ud},
indicating that supersymmetry is not sufficient to discard these
causally pathological solutions. Moreover, a particular issue in
the dual description of gravity theories by gauge theories is the
conjecture linking closed time-like curves on the gravity side and
non-unitarity on the gauge side
\cite{Herdeiro:2000ap,Caldarelli:2004mz}. It was indeed shown
\cite{Herdeiro:2000ap} in the context of BMPV black holes
\cite{Breckenridge:1996is} that the regime of parameters in which
there exists naked closed timelike curves is also the regime in
which unitarity is violated in the dual CFT. Also, half BPS
excitations in $AdS_5\times S^5$ in IIB sugra can be mapped to
fermions configurations \cite{Lin:2004nb}. Causality violation is
shown to be related to Pauli exclusion principle in the dual
theory \cite{Caldarelli:2004mz}.

It was recently shown that a new class of solutions in
five-dimensional supergravity \cite{Gimon:2003ms} could be viewed
as Kerr black holes embedded in a G\"odel universe. Their
peculiarity lies principally in the presence of closed time-like
curves in the large-radius asymptotic region so that these
solutions combine the two aforementioned puzzles. Due to their
unusual asymptotic behavior, being neither flat nor anti-de
Sitter, the traditional methods for computing the conserved
quantities associated with such black holes generically fail
\cite{Klemm:2004wq}. However, as shown in \cite{Barnich:2005kq},
their mass, angular momenta and electric charge could still
consistently be defined, and were shown to satisfy both
generalized Smarr formula and first law of black holes mechanics.
This suggests that this class of black holes could also have an
interpretation as thermodynamical objects, whose properties are
worth further investigations.

A road which has revealed successful in bringing insights into
properties of quantum gravity is the study of lower-dimensional
systems (see e.g. \cite{Carlip:1998uc}). A particularly eloquent
example is that of the BTZ black hole, solution of 2+1 gravity
with a negative cosmological constant
\cite{Banados:1992wn,Banados:1992gq}. Since its discovery, it has
appeared as a useful toy model to address namely the problem of
black hole entropy in a simpler setting (for reviews, see e.g.
\cite{Carlip:2005zn,Carlip:1998qw,Banados:1998sm}). In particular,
Strominger's derivation of BTZ black holes' entropy exactly
reproduces their geometrical Bekenstein-Hawking entropy
\cite{Strominger:1997eq}. This computation essentially relied on
two earlier works: one by Brown and Henneaux \cite{Brown:1986nw},
which showed that the canonical realization of asymptotic
symmetries of $AdS_3$ is represented by two Virasoro algebras with
non-vanishing central charge, and another by Cardy et al. via the
so-called Cardy formula
\cite{Bloete:1986qm,Cardy:1986ie,Dijkgraaf:2000fq}, which allows
to count in the semi-classical limit the asymptotic density of
states of a conformal field theory, even if the full details of
the theory are not known. Application of the latter with the
former central charge strikingly yields the expected number of
states, even if a deep explanation of ``why it works" so well is
still missing so far (for a discussion on the application of the
Cardy formula, see e.g. \cite{Carlip:2005zn}).

In this note, we would like to use a similar philosophy to grasp
with some properties of the aforementioned G\"odel black holes
through the canonical representation of their asymptotic
symmetries. The theory of interest here will be (2+1)-dimensional
Einstein-Maxwell-Chern-Simons theory, which can be viewed as a
lower-dimensional toy-model for the bosonic part of $D=5$
supergravity, since the field content and the couplings of both
theories are similar. This theory was shown to admit as solutions
G\"odel universes and a three-dimensional version of G\"odel black
holes, displaying the same peculiar properties as their
higher-dimensional counterparts \cite{Banados:2005da}.

Our analysis will present analogies with the one performed in
$AdS_3$ space since there is a close relationship between $AdS_3$
and $3d$ G\"odel space. Indeed, the latter can be seen as a
squashed $AdS_3$ space, where the original isometry group is
broken from $\SL \times \SL$ to $\SL \times U(1)$, as pointed out
in \cite{Rooman:1998xf}. In the context of string theory, the $3d$
G\"odel metric was shown to be part of the target space of an
exact two-dimensional CFT, obtained as an asymmetric marginal
deformation of the $\SL$ WZW model \cite{Israel:2003cx}. In this
case, the effect of the deformation amounts to break the original
$\widehat{\SL} \times \widehat{\SL}$ symmetry of the model down to
$\widehat{\SL} \times \widehat{U(1)}$. As we will show, a similar
pattern will appear at the level of asymptotic symmetries.

After having briefly recalled in Sect.~2 our general setup, we
will compute in Sect.~3 the asymptotic symmetry algebra of G\"odel
spaces. We then define, in Sect.~4, a class of field
configurations, which we will refer to as asymptotically G\"odel
space-times in three dimensions, encompassing the previously
mentioned black holes solutions. In Sect.~5, we represent the
algebra of charges by covariant Poisson brackets and show that the
asymptotic symmetry algebra admits central extensions. We conclude
in Sect.~6 by discussing some of the results, arguing that part of
the entropy of the G\"odel black holes can be obtained via the
Cardy formula.

\section*{Note added in v2}

In version 1 of this paper, it has been incorrectly claimed that the obtained Virasoro central charge was negative. This reasoning was a consequence of a sign mistake in \cite{Barnich:2006av} that has been corrected in the revised version \cite{Barnich:2006av4}. All mentions and analyses of this minus sign have been removed in the current version.

\section{General setup}

Let us start with the Einstein-Maxwell-Chern Simons theory in $2+1$
dimensions,
\begin{equation}
I = \frac{1}{16 \pi G} \int \, d^3x \,\left[ \sqrt{-g}\left( R +
\frac{2}{l^2} - \frac{1}{4} F^2 \right) -
\frac{\alpha}{2}\eps^{\mu\nu\rho} A_\mu F_{\nu\rho}
\right].\label{action}
\end{equation}
The gauge parameters of the theory $(\xi,\lambda)$, where $\xi$ generates infinitesimal diffeomorphisms and $\lambda$ is the parameter of $U(1)$ gauge transformations are endowed with the Lie algebra structure
\begin{equation}
[(\xi,\lambda),(\xi^\prime,\lambda^\prime)]_{G} =
([\xi,\xi^\prime],[\lambda,\lambda^\prime]),\label{eq:Lie}
\end{equation}
where the $[\xi,\xi^\prime]$ is the Lie bracket and
$[\lambda,\lambda^\prime]\, \equiv \cL_\xi \lambda^\prime -
\cL_{\xi^\prime}\lambda$. We will denote for compactness the fields as $\phi^i\equiv (g_{\mu\nu},A_\mu)$ and the gauge parameters as $f^\alpha = (\xi^\mu,\lambda)$. For a given field $\phi$, the gauge parameters $f$ satisfying
\begin{equation}
\cL_\xi  g_{\mu\nu} \approx 0, \qquad \cL_\xi
A_{\mu} + \d_\mu \lambda \approx 0, \label{eq:red}
\end{equation}
where $\approx$ is the on-shell equality, will be called the exact symmetry parameters of $\phi$. Parameters $(\xi,\lambda) \approx 0$ are called trivial symmetry parameters.

In order to study the conserved charges for this theory, one canonically constructs the following $1$-form in spacetime \cite{Barnich:1994db,Barnich:2001jy},
\begin{equation}
k_{(\xi,\lambda)}[\dv \phi ; \phi]= k^{exact}_{(\xi,\lambda)}[\dv
\phi ; \phi] - k^{s}_{(\cL_\xi g,\cL_\xi A + d\lambda)}[\dv
\phi,\phi],\label{charge_k}
\end{equation}
which is associated both with an arbitrary variation $\dv \phi$
around $\phi$ and with parameters $(\xi,\lambda)$. The form of
$k^{exact}$ was given in equation~(65) of \cite{Banados:2005da}
and will not be reproduced here. The supplementary term
\begin{equation}
k^{s}_{(\cL_\xi g,\cL_\xi A + d\lambda)}[\dv \phi,\phi] =
\frac{\sqrt{-g}}{32\pi G}\left(\dv g_{\mu \alpha}(D^\alpha \xi_\nu
+ D_\nu \xi^\alpha)-\dv A_\mu (\cL_\xi A_\nu + \d_\nu \lambda)
\right) \epsilon^{\mu\nu}_{\,\,\,\,\,\alpha}dx^\alpha
\end{equation}
vanishes for exact symmetries but is relevant in the asymptotic context.

The form $k_{(\xi,\lambda)}[\dv \phi ; \phi]$ enjoys the following
properties
\begin{itemize}
\item For $\phi$ a solution of the equations of motion and $\dv \phi$ a solution of the linearized equations of motion around $\phi$, there is a one-to-one correspondence between the non-trivial exact symmetry parameters of $\phi$ and the conserved $1$-forms $d k \approx 0$ of the linearized theory which are non-trivial, i.e. not exact on-shell. These forms are given by $k_{(\xi,\lambda)}=k^{exact}_{(\xi,\lambda)}$ modulo the addition of exact forms or on-shell vanishing forms. The integral of $k$ on the circle $t=constant$, $r=constant$ then yields the finite and conserved quantity associated to $(\xi,\lambda)$ that depends only on the homology class of the circle.
\item The form $k$ is constructed out of the equations of motion. It therefore does not depend on boundary terms that may be added to the lagrangian.
\end{itemize}
Additional properties of $k$ are discussed in \cite{Barnich:2004uw,Barnich:2006av}. The set of fields $\phi$, $\dv \phi$ and gauge parameters $(\xi,\lambda)$ that satisfy the conditions
\begin{eqnarray}
\oint_{S^\infty} \dv k_{(\xi,\lambda)}[\dv \phi,\phi] = 0, \label{cond_int} \\
\oint_{S^\infty} k^{s}_{(\cL_\xi g,\cL_\xi A + d\lambda)}[\dv \phi,\phi] = 0,\label{cond_alg}
\end{eqnarray}
where $S^\infty$ is the circle $t=constant$, $r=constant
\rightarrow \infty$ define a space of fields and parameters which
we denote as the integrable space $\cI$. In this space, we define
the charges difference between the fields $\bar \phi$ and $\phi$
associated with $(\xi,\lambda)$ as
\begin{equation}
\cQ_{(\xi,\lambda)}[\phi,\bar \phi] = \oint_{S^\infty} \int_\gamma k_{(\xi,\lambda)}[\dv \phi,\phi] + \cN_{(\xi,\lambda)}[\bar \phi],
\end{equation}
where $\gamma$ is a path in field space contained in $\cI$ and $\cN_{(\xi,\lambda)}[\bar \phi]$ is an arbitrary normalization constant. Condition~\eqref{cond_int} ensures that the charge is independent on smooth deformations of the path $\gamma$.

Let us introduce for later convenience a subset of $\cI$ with
elements $(\phi,f[\phi])$ such that for each field $\phi$ the set
of parameters $f[\phi]$ form a closed Lie algebra under the
bracket defined in \eqref{eq:Lie} and such that all these algebras
are isomorphic. Let denote this algebra by $\cA$. Using the
conditions~\eqref{cond_int}-\eqref{cond_alg}, one can then show
\cite{Barnich:2007bientot} that for any solutions $\bar \phi$ and
$\phi$ in the integrable space, and for any $(\xi,\lambda)$,
$(\xi^\prime, \lambda^\prime)$ in $\cA$, the expression\footnote{Note added in version 2: The order of indices on the right-hand side of the expression have been changed with respect to version 1.}
\begin{eqnarray}
\cK_{(\xi,\lambda),(\xi^\prime,\lambda^\prime)}[\bar \phi] = \int_{S^\infty} k_{(\xi, \lambda)}[(\cL_{\xi^\prime} \bar g_{\mu\nu},\cL_{\xi^\prime} \bar A_{\mu} +
\d_\mu \lambda^\prime);(\bar g,\bar A)] \label{eq:cc}
\end{eqnarray}
is a Chevalley-Eilenberg 2-cocycle on the Lie algebra $\cA$ and the Poisson bracket defined by
\begin{equation}
\left\{ \cQ_{(\xi,\lambda)}[\phi,\bar \phi],
\cQ_{(\xi^\prime,\lambda^\prime)}[\phi,\bar \phi] \right\} \equiv
\oint_{S^\infty} k_{(\xi^\prime, \lambda^\prime)}[(\cL_\xi
g_{\mu\nu},\cL_\xi A_{\mu} + \d_\mu \lambda);(g, A)]
\end{equation}
obeys
\begin{equation}
\left\{ \cQ_{(\xi,\lambda)}[\phi,\bar \phi],
\cQ_{(\xi^\prime,\lambda^\prime)}[\phi,\bar \phi] \right\} =
\cQ_{[(\xi,\lambda),(\xi^\prime,\lambda^\prime)]_{G}}[\phi,\bar \phi] -
\cN_{[(\xi,\lambda),(\xi^\prime,\lambda^\prime)]_{G}}[\bar g,\bar
A] + \cK_{(\xi,\lambda),(\xi^\prime,\lambda^\prime)}.\label{formula}
\end{equation}
The central extension~\eqref{eq:cc} is considered as trivial if it
can be reabsorbed in the normalization of the charges. Saying it
differently, a central charge is non-trivial if it cannot be
written as a function of the bracket
$[(\xi,\lambda),(\xi^\prime,\lambda^\prime)]_{G}$ only.

Given a solution $\bar \phi$, one can define an algebra $\cA$ of asymptotic symmetries $(\xi,\lambda)$ by the following conditions,
\begin{itemize}
\item We define the parameters $(\xi,\lambda)$ such that the
leading order of the expressions $\cL_\xi \bar  g_{\mu\nu}$ and $\cL_\xi \bar A_{\mu} + \d_\mu \lambda$ close to the boundary $S^\infty$ vanishes.
\item We require the expression $\cK_{(\xi,\lambda),(\xi^\prime,\lambda^\prime)}[\bar \phi]$ to be a finite constant.
\item We impose that the Lie bracket of two such parameters also satisfies the latter conditions.
\end{itemize}
The first condition is an adaptation of the exact symmetry
equations~\eqref{eq:red} in the asymptotic context. In pure
gravity, asymptotic Killing vectors can be defined similarly as
vectors fields obeying the Killing equations to ``as good an
approximation as possible'' as one goes to the
boundary~\cite{Waldbook}. The second condition expresses
finiteness and conservation of the form~\eqref{charge_k}
integrated over $S^\infty$ and evaluated on the background in the
particular case where $\dv\phi$ is $(\cL_{\xi^\prime} \bar
g_{\mu\nu},\cL_{\xi^\prime} \bar A_{\mu} + \d_\mu
\lambda^\prime)$. Because we will require that the phase space be
left invariant under the asymptotic symmetry algebra (see section
4), such a $\dv \phi$ will be tangent to the phase space. As will
be required for any tangent vector to the phase space, such a $\dv
\phi$ has to be associated with finite and conserved charges (see
also section 4). In fact, the second condition are the constraints
on finiteness and conservation that one can impose already at this
stage. The third condition simply ensures that the asymptotic
symmetry parameters form a Lie algebra.

In what follows, we use this definition to compute the asymptotic
symmetries for the G\"odel spacetime and then construct a space of
fields $\cF$ consistent with $\cA$ and satisfying the
conditions~\eqref{cond_int}-\eqref{cond_alg}.

\section{G\"odel asymptotic symmetry algebra}
\label{sec_ask}

It was shown in \cite{Banados:2005da} that the equations of motion derived from~\eqref{action} admit the solution
\begin{eqnarray}\label{GodelMetric}
\bar{ds}^2  &=& \eps dt^2  - 4\alpha r d t d\varphi + (2r -
\frac{2}{l^2}|1-\alpha^2l^2| r^2)d \varphi^2 +
\frac{1}{-2 \eps r+ \Upsilon^{-1}r^2}dr^2 \label{Sol_backgr} \\
\bar A &=&  \frac{2}{l}\sqrt{|1-\alpha^2l^2|}\, r d\varphi, \nonumber
\end{eqnarray}
where $\eps = \text{sgn}(1-\alpha^2 l^2)$, $\Upsilon = \frac{l^2}{2(1+\alpha^2l^2)}$ and $\varphi \in [0,2\pi]$. For $\eps = -1$, this solution is the $3d$ part of the two parameter generalization \cite{Reboucas:1983hn} of the G\"odel spacetime \cite{Godel:1949ga} where the stress-energy tensor of the perfect fluid supporting the metric is generated by the gauge field. For $\epsilon = +1$, the solution will be called the tachyonic G\"odel spacetime because the perfect fluid supporting the metric is tachyonic. We will use this solution as background in the two sectors of the theory $\eps = \pm 1$.

For $\eps = -1$, the G\"odel solution (\ref{GodelMetric}) admits 5 non-trivial exact symmetries $(\xi,\lambda)$,
\begin{eqnarray}
(\xi_{(1)},0) &=& (\d_t,0) ,\nonumber  \\
(\xi_{(2)},0) &=& (2\alpha \Upsilon \d_t +\d_\varphi,0), \nonumber\\
(\xi_{(3)},0) &=&(
\frac{2\alpha\Upsilon}{\sqrt{1+2\Upsilon/r}}\sin\varphi \d_t -
\sqrt{2 \Upsilon r+r^2}\cos\varphi \d_r +
\frac{r+\Upsilon}{\sqrt{2\Upsilon r+r^2}}\sin\varphi \d_\varphi,0),
 \label{ask_exact}\\
(\xi_{(4)},0)
&=&(\frac{2\alpha\Upsilon}{\sqrt{1+2\Upsilon/r}}\cos\varphi \d_t +
\sqrt{2 \Upsilon r+r^2}\sin\varphi \d_r +
\frac{r+\Upsilon}{\sqrt{2\Upsilon r+r^2}}\cos\varphi
\d_\varphi,0),\nonumber\\
(0,\lambda_{(1)}) &=& (0,1).\nonumber
\end{eqnarray}
The four Killing vectors form a $ \mathbb{R} \oplus so(2,1)$ algebra. In the case $\eps = +1$, only the two first vectors are Killing vectors.

Let us now compute the asymptotic symmetries of this background solution $\bar \phi$. They are of the form
\begin{eqnarray}
\xi &=& \chi_\xi(r)\tilde \xi(t,\varphi)+o(\chi_\xi(r))\nonumber \\
\lambda &=& \chi_\lambda(r)\tilde \lambda(t,\varphi)+o(\chi_\lambda(r)),\label{form_aKv}
\end{eqnarray}
for some fall-offs $\chi_\xi(r)$, $\chi_\lambda(r)$ and functions $\tilde \xi(t,\varphi)$, $\tilde \lambda(t,\varphi)$ to be determined. For such parameters, one has
\begin{equation}
\cL_\xi \bar  g_{\mu\nu} = O(\rho_{\mu\nu}), \qquad \cL_\xi \bar A_{\mu} + \d_\mu \lambda = O(\rho_\mu), \label{cond_eps}
\end{equation}
where $\rho_{\mu\nu}$ and $\rho_\mu$ depend on the explicit form of the parameters~\eqref{form_aKv}. Equations~\eqref{cond_eps} are satisfied to the leading order in $r$ when one imposes $\cL_\xi \bar  g_{\mu\nu} = o(\rho_{\mu\nu})$ and $\cL_\xi \bar A_{\mu} + \d_\mu \lambda = o(\rho_\mu)$. If one solves these equations with the highest order in $r$ for $\chi_\xi(r)$ and $\chi_\lambda(r)$, one gets the unique solution
\begin{eqnarray}
\xi &=& (F(t,\varphi) +o(r^0)) \d_t +( - r \d_\varphi
\Phi(\varphi)
+o(r^1))\d_r + (\Phi(\varphi)+o(r^0) ) \d_\varphi,\label{asK} \\
\lambda &=& \lambda(t,\varphi) + o(r^0),
\end{eqnarray}
where $F(t,\varphi)$ and $\Phi(\varphi)$ are arbitrary functions. We now require the central extension~\eqref{eq:cc} to be a finite constant. The term diverging in $r$ in~\eqref{eq:cc} vanishes if we impose
$\xi^\varphi = \Phi(\varphi)+o(r^{-1})$. The central extension is then constant by requiring
\begin{equation}
F(t,\varphi) = F(\varphi),\qquad \lambda(t,\varphi)= \lambda(\varphi).
\end{equation}
The resulting expression for~\eqref{eq:cc} is given by
\begin{eqnarray}
\hspace{-6pt}K_{f^\prime,f}[\bar \phi]\hspace{-3pt}
&=&\hspace{-3pt} \frac{1}{16\pi G}\hspace{-3pt}
\int_0^{2\pi}\hspace{-7pt} d\varphi \Big[ 2\alpha \Upsilon
\d_\varphi \Phi^\prime \d_\varphi^2\Phi
\hspace{-1pt}-\hspace{-1pt}\frac{\eps}{2\alpha\Upsilon}\d_\varphi
F F^\prime \hspace{-1pt}+\hspace{-1pt} 2\eps \Phi^\prime
\d_\varphi F\hspace{-1pt}+ \hspace{-1pt}\alpha\d_\varphi \lambda
\lambda^\prime - (f \hspace{-2pt}\leftrightarrow
\hspace{-2pt}f^\prime) \Big].\label{eq:cc2}
\end{eqnarray}
The asymptotic symmetries just found form a subalgebra $\cA$ of
the bracket~\eqref{eq:Lie}. The asymptotic symmetries which are of
the form
\begin{equation}
\xi = o(r^{0})\d_t +o(r^1)\d_r + o(r^{-1})\d_\varphi, \qquad
\lambda =o(r^{0}),\label{triv}
\end{equation}
will be considered as trivial because (i) they form an ideal of the algebra $\cA$, (ii) for any $f$ of the form~\eqref{triv} and $f^\prime \in \cA$, the associated central charge $K_{f,f^\prime}[\bar \phi]$ vanishes. We define the asymptotic symmetry algebra $\mathfrak{Godel_3}$ as the quotient of $\cA$ by the trivial asymptotic symmetries~\eqref{triv}. This algebra can thus be expressed only in terms of the leading order functions $F(\varphi)$, $\Phi(\varphi)$ and $\lambda(\varphi)$. By setting $\hat f=[f,f^\prime]_G$, one can write the $\mathfrak{Godel_3}$ algebra explicitly as
\begin{equation}
\hat F(\varphi) =  \Phi \d_\varphi F^\prime - \Phi^\prime
\d_\varphi F, \qquad \hat \Phi(\varphi) = \Phi \d_\varphi
\Phi^\prime - \Phi^\prime \d_\varphi \Phi,\qquad  \hat \lambda =
\Phi \d_\varphi \lambda^\prime - \Phi^\prime \d_\varphi \lambda.\label{eq:algGo}
\end{equation}
A convenient basis for non-trivial asymptotic symmetries consists in the following generators
\begin{eqnarray}
l_n &=& \{ (\xi,\lambda) \in \cA | F(\varphi) = 2\alpha \Upsilon e^{i n \varphi}, \, \Phi(\varphi) = e^{i n \varphi}, \, \lambda(\varphi) = 0 \}, \nonumber\\
t_n &=& \{ (\xi,\lambda) \in \cA | F(\varphi) = e^{i n \varphi}, \, \Phi(\varphi) = 0, \, \lambda(\varphi) = 0 \},\label{gen_back}\\
j_n &=& \{ (\xi,\lambda) \in \cA | F(\varphi) = 0, \,
\Phi(\varphi) = 0, \, \lambda(\varphi) = e^{i n \varphi} \}.
\nonumber
\end{eqnarray}
In terms of these generators, the $\mathfrak{Godel_3}$ algebra
reads
\begin{eqnarray}
i[ {l_m},{l_n}]_G &= &(m-n) l_{m+n},\nonumber\\
i[ {l_m},{t_n} ]_G &=& -n t_{m+n},\label{alg_Godel}\\
i[{l_m},{j_n} ]_G & =& -n j_{m+n},\nonumber
\end{eqnarray}
while the other commutators are vanishing. One can recognize the exact symmetry parameters~\eqref{ask_exact} as a subalgebra of $\mathfrak{Godel_3}$. Indeed, one has $t_0 \sim (\xi_{(1)},0)$, $l_0 \sim (\xi_{(2)},0)$, $l_{-1} \sim (-i\xi_{(3)}+\xi_{(4)},0)$, $l_1 \sim (i \xi_{(3)}+\xi_{(4)},0)$ and $j_0 \sim (0,\lambda_{(1)})$ where $\sim$ denote the belonging to the same equivalence class of asymptotic symmetries.

In $AdS_3$, the exact $so(2,1)\oplus so(2,1)$ algebra is enhanced
in the asymptotic context to two copies of the Witt algebra. The
G\"odel metric can be interpreted as a squashed $AdS_3$ geometry,
which breaks the original $so(2,1) \oplus so(2,1)$ symmetry
algebra down to $u(1) \oplus so(2,1) $ \cite{Rooman:1998xf}. The
exact Killing symmetry algebra is here enhanced to a semi-direct
sum of a Witt algebra with a $\widehat{u(1)}$ loop algebra.
Moreover, the gauge sector $u(1)$ is enhanced to another
$\widehat{u(1)}$ loop algebra also forming an ideal of the
$\mathfrak{Godel_3}$ algebra.

\section{Asymptotically G\"odel fields}

We defined in the previous section the asymptotic symmetry algebra
$\mathfrak{Godel_3}$ by a well-defined procedure starting from the
background $\bar \phi$. One can ask which are the field
configurations $\phi$ such that the preceding analysis leads to
the same algebra~\eqref{eq:algGo} with $\bar \phi$ replaced by
$\phi$. The subset of such field configurations which is preserved
under the action of the asymptotic symmetry algebra will then
provide a natural definition of asymptotically G\"odel fields
$\cF$. A set of fields satisfying these conditions is given by
\begin{eqnarray}
g_{tt} &=& \eps+r^{-1}\overset{(1)}{g_{tt}}+O(r^{-2}), \qquad
g_{tr} = O(r^{-2}), \qquad g_{t\varphi}=-2\alpha  r +
\overset{(1)}{g_{t\varphi}}+O(r^{-1}) ,\nonumber\\
g_{rr} &=&
\frac{\Upsilon}{r^2}+r^{-3}\overset{(1)}{g_{rr}}+O(r^{-4}), \qquad
g_{r\varphi}=r^{-1}\overset{(1)}{g_{r\varphi}}+O(r^{-2}),\nonumber\\
&& \qquad g_{\varphi\varphi} =
-\frac{2}{l^2}|1-\alpha^2l^2|r^2+r^{1}\overset{(1)}{g_{\varphi\varphi}}+O(r^{0}),\label{BC1}\\
A_t &=& -\frac{\sqrt{(1-\alpha^2l^2)\eps}}{\alpha l
}+r^{-1}\overset{(1)}{A_{t}}+O(r^{-2}),\qquad A_r =
r^{-2}\overset{(1)}{A_r}+O(r^{-3}), \nonumber\\
&&\qquad A_\varphi =
\frac{2}{l}\sqrt{\abs{1-\alpha^2l^2}}r+\overset{(1)}{A_{\varphi}}+O(r^{-1}),\nonumber
\end{eqnarray}
where all functions $\overset{(1)}{g_{tt}}$, \dots$\,$ depend
arbitrarily on $t$ and $\varphi$. In order for these field
configurations be left invariant under the asymptotic symmetries,
one has furthermore to restrict the subleading component of
$\xi^\varphi$ to $\xi^\varphi = \Phi(\varphi)+O(r^{-2})$. The
asymptotic symmetries thus become
\begin{eqnarray}
\xi &=& (F(\varphi) +o(r^0)) \d_t +( - r \d_\varphi
\Phi(\varphi)
+o(r^1))\d_r + (\Phi(\varphi)+O(r^{-2}) ) \d_\varphi, \\
\lambda &=& \lambda(\varphi) + o(r^{0}),
\end{eqnarray}
and always contain the asymptotic form of the exact symmetries
\eqref{ask_exact}.

However, for the purpose of providing a well-defined
representation of the asymptotic symmetry algebra, one has to
restrict the definition of fields $\cF$ by selecting those
satisfying \eqref{cond_int}-\eqref{cond_alg} and admitting finite
and conserved charges. These conditions are met if the following
differential equation hold,
\begin{equation}
\overset{(1)}{g_{\varphi\varphi}}-\eps
\Upsilon^{-2}\overset{(1)}{g_{rr}}+4\alpha \eps
\overset{(1)}{g_{t\varphi}}+\frac{2(\alpha^2l^2-1)}{l^2}\overset{(1)}{g_{tt}}+\frac{\eps}{\alpha
\Upsilon}\d_t \overset{(1)}{g_{r\varphi}}+\frac{2\eps \sqrt{\eps
(1-\alpha^2l^2)}}{\alpha l \Upsilon}(\d_t
\overset{(1)}{A_r}+\overset{(1)}{A_t}) = 0.\label{BC2}
\end{equation}
We finally define the set of asymptotically G\"odel fields $\phi =
(g,A)$ as those satisfying the boundary conditions~\eqref{BC1} and
\eqref{BC2}.

In general, the asymptotic symmetries are allowed to depend arbitrarily on the fields, $(\xi[g,A],\lambda[g,A])$. They should however, by construction,
 obey the same algebra $\mathfrak{Godel_3}$. A basis for the asymptotic symmetries of $\phi$ can be written as
 \begin{eqnarray}
l_n &=& \{ (\xi,\lambda) \in \cA | F(\varphi) = 2\alpha \Upsilon N_l[g,A] e^{i n \varphi}, \, \Phi(\varphi) = e^{i n \varphi}, \, \lambda(\varphi) = 0 \}, \nonumber\\
t_n &=& \{ (\xi,\lambda) \in \cA | F(\varphi) = N_t[g,A] e^{i n \varphi}, \, \Phi(\varphi) = 0, \, \lambda(\varphi) = 0 \},\label{gen_gen}\\
j_n &=& \{ (\xi,\lambda) \in \cA | F(\varphi) = 0, \, \Phi(\varphi) = 0, \, \lambda(\varphi) =N_j[g,A]  e^{i n \varphi} \}, \nonumber
\end{eqnarray}
The additional solution-dependent normalizations are constrained by $N_l[\bar g,\bar A] = N_t[\bar g,\bar A]= N_j[\bar g,\bar A]= 1$ in order to match  the asymptotic symmetries defined for the background.

Note that besides the background itself the asymptotically G\"odel
fields contain the three parameters ($\nu$, $J$, $Q$) particle
$(\eps= -1)$ and black hole $(\eps = +1)$ solutions found
in~\cite{Banados:2005da}, \footnote{The solutions written in
eq.~(49)-(57) of~\cite{Banados:2005da} differs with the solutions
written here by the change of coordinates $r^{here} =
\frac{r^{there}}{\sqrt{|8G\mu^{there}|}}$, $ t^{here} =
\sqrt{|8G\mu^{there}|} t^{there}$, $\nu = 2 \eps
\sqrt{|8G\mu^{there}|}$.}
\begin{eqnarray}
ds^2  &=&\eps d t^2  - 4\alpha r d t d\varphi +
(-\frac{4GJ}{\alpha}+ 8G\eps \nu r -
\frac{2}{l^2}|1-\alpha^2l^2| r^2)d \varphi^2
+ \left( \frac{4GJ}{\alpha}\eps - 8G\nu r
+ \frac{r^2}{\Upsilon} \right)^{-1} dr^2 \nonumber\\
A &=&   -\eps\frac{\sqrt{|1-\alpha^2l^2|}}{\alpha l}  d t +(
-\frac{4G}{\alpha} Q + \frac{2}{l}\sqrt{(1-\alpha^2l^2)\eps}\, r)
d\varphi. \label{Sol_gen}
\end{eqnarray}

\section{Poisson algebra}

We are now ready to represent the asymptotic algebra
$\mathfrak{Godel_3}$ by associated charges in the space of
configurations defined in~\eqref{BC1}-\eqref{BC2}. An explicit
computation shows that the charges associated with each
generator~\eqref{gen_gen} are in general non-vanishing. We denote
these charges by $L_n \equiv \cQ_{l_n}[\phi,\bar \phi]$, $T_n
\equiv \cQ_{t_n}[\phi,\bar \phi]$ and $J_n \equiv
\cQ_{j_n}[\phi,\bar \phi]$. On the contrary, all trivial
asymptotic symmetries are associated with vanishing charges as it
should. This provides additional justification for the quotient
$\mathfrak{Godel_3}$ taken in section~\ref{sec_ask}.

The central extensions~\eqref{eq:cc2} may be explicitly computed
for any pair of generators of the background~\eqref{gen_back}. The
only non-vanishing terms are
\begin{eqnarray}
i K_{l_m,l_n} &=& \frac{c}{12} m(m^2+\eps)\delta_{n+m},\nonumber\\
i K_{t_m,t_n} &=& -\frac{\eps}{8G \alpha \Upsilon} m\delta_{m+n,0}.\label{final_cc}\\
i K_{j_m,j_n} &=& \frac{\alpha}{4G} m \delta_{m+n}.\nonumber
\end{eqnarray}
where the Virasoro-type central charge $c$ reads
\begin{equation} c = \frac{6\alpha
\Upsilon}{G}= \frac{3\alpha l^2}{(1+\alpha^2l^2)G}.
\end{equation}
According to~\eqref{formula}, the G\"odel algebra is finally represented at the level of charges
by the following centrally extended Poisson algebra
\begin{eqnarray}
i\{L_m,L_n\} &=& (m-n)(L_{m+n}-\cN_{l_{m+n}}) + \frac{c}{12}m(m^2+\eps)\delta_{m+n},\nonumber\\
i\{L_m,T_n\} &=& -n (T_{m+n}-\cN_{t_{m+n}}), \nonumber\\
i\{T_m,T_n\} &=& -\frac{\eps}{8G \alpha \Upsilon} m\delta_{m+n},\label{C1} \\
i\{L_{m},J_{n}\} &=& -n(J_{m+n}-\cN_{j_{m+n}}),\nonumber\\
i\{J_{m},J_{n}\} &=& \frac{\alpha}{4G}m \delta_{m+n}.\nonumber
\end{eqnarray}
The central extensions~\eqref{final_cc} are non-trivial because
they cannot be absorbed into the (undetermined classically)
normalizations of the generators. The $L_n$ form a Virasoro
algebra while the two loop algebras $\{ t_n \}$, $\{ j_n \}$ are
represented by centrally extended $\widehat{u(1)}$ affine
algebras.

\section{Discussion}

In $3d$ asymptotically anti-de Sitter spacetimes, the asymptotic
symmetry algebra which consists of two copies of the Virasoro
algebra \cite{Brown:1986nw} allows one to compute the entropy of
the BTZ black hole via the Cardy formula \cite{Strominger:1997eq}.
One may wonder if an analogous derivation based on the asymptotic
algebra~\eqref{C1} could be performed.

It turns out that the analysis in G\"odel spacetimes is more
tricky. The G\"odel black holes are given in~\eqref{Sol_gen} when
$\eps = +1$. In this case, the $r$ coordinate has the range
$-\infty < r < \infty$. The solution~\eqref{Sol_gen} displays an
horizon and therefore describes a regular black hole only if the
inequality
\begin{equation}
2G \nu^2  \geq \frac{J}{2\alpha \Upsilon}\label{eq:bounds}
\end{equation}
holds. The tachyonic G\"odel solution corresponds to $\nu =
+\frac{1}{4G}$, $J=Q=0$. Because the solutions with $\nu$, $J$ and
$Q$ are related by the change of coordinates $r \rightarrow -r$,
$\varphi \rightarrow -\varphi$ with the solutions $-\nu$, $J$,
$-Q$, the conserved quantity $\nu$ associated to $\d_t$ does not
provide a satisfactory definition of mass. However, one can define
the quantity $\mu = 2\eps G\nu^2$ which is by definition positive
for black holes and which equals $-\frac{1}{8G}$ for the G\"odel
background. In particular, in the anti-de Sitter limit $\alpha^2
l^2 \rightarrow 1$, $\mu$ correctly reproduces the mass gap
between the zero mass BTZ black hole and anti-de Sitter space. It
was shown in \cite{Banados:2005da} that this quantity is
associated with the Killing vector $4G \eps \nu \d_t$. Note also
that $\d_\varphi$ is associated with $-J + \frac{Q^2}{4\alpha}$.

Choosing the normalization $N_{l} = 4G \nu$, the charge associated
with the generator $l_0$ of~\eqref{gen_gen} becomes for the black
holes
\begin{equation}
L_0 = 2\alpha \Upsilon \mu - J + \frac{Q^2}{4\alpha} - \frac{\alpha \Upsilon}{4G} + Q_{l_0}[\bar \phi].
\end{equation}
When $\alpha > 0$, the inequality~\eqref{eq:bounds} imposes that the spectrum of $L_0$ is bounded from below. The Virasoro generators $L_n$ may then be associated with operators acting on a ground state with minimal $L_0$-eigenvalue. When $\alpha <0$, one may instead consider the generators $L^\prime_n = - L_{-n}$ satisfying also a Virasoro algebra
\begin{eqnarray}
i\{L^\prime_m,L^\prime_n\} &=& (m-n)(L^\prime_{m+n}-\cN_{l^\prime_{m+n}}) + \frac{c^\prime}{12}m(m^2+1)\delta_{m+n},
\end{eqnarray}
with $c^\prime = -c = -6\alpha \Upsilon/G$ and for which $ L^\prime_0 = -L_0$ is also bounded from below. Remark that in any of these two cases, the classical Virasoro central charge ($c$ for $\alpha >0$ and $c^\prime$ for $\alpha<0$) is positive while the Virasoro zero-mode ($L_0$ for $\alpha > 0$ and $L^\prime_0$ for $\alpha < 0$) is bounded from below. This behavior of the Virasoro charges is not pathological even with the presence of the G\"odel close timelike curves. In the anti-de Sitter limit $\alpha^2 \rightarrow 1/l^2$, the central charge tends to the usual AdS$_3$ central charge $3l/2G$.

The Bekenstein-Hawking entropy associated with the black hole solutions~\eqref{Sol_gen} is given by
\begin{equation}
S_{BH} = 2\pi \sqrt{\alpha \Upsilon G^{-1}(2\alpha \Upsilon \mu-
J)}+2\pi \sqrt{2\alpha^2\Upsilon^2 G^{-1}\mu}.\label{BH}
\end{equation}
Let us consider without loss of generality the case $\alpha >0$
and define $\Delta_0$ as the value of $L_0$ for the zero mass
black hole $\mu=J=Q=0$, $\Delta_0 = -\alpha
\Upsilon/(4G)+\cQ_{l_0}[\bar \phi]$. We observe that the first
term in~\eqref{BH} may be written as $2\pi
\sqrt{(c-24\Delta_0)L_0/6}$ for $\Delta_0 = 0$ or $\Delta_0 =
-\alpha \Upsilon/(2G)$ and for $Q=0$ in the large mass $\mu \gg
1/(8G)$ limit. In the semi-classical limit $\alpha \Upsilon \gg
G$, the latter formula is the Cardy formula 
\cite{Bloete:1986qm,Cardy:1986ie,Dijkgraaf:2000fq} for the
Virasoro algebra with generators $L_n$.

It is possible to reproduce the second part of the
entropy~\eqref{BH} via the Cardy formula by introducing operators
$\hat T_n$ to each element of the affine algebra $T_n$, applying
the Sugawara procedure to obtain a new Virasoro algebra $\tilde
L_n$ with central charge $\tilde c = 1$ and by appropriately
choosing the lowest value $\tilde \Delta_0$ of $\tilde L_0$. In
this case, the effective central charge $\tilde c-24\tilde
\Delta_0 = 6\alpha \Upsilon/G$ equals the effective central
charge $c - 24 \Delta_0$ in the initial Virasoro sector.
However, this construction \emph{a posteriori} is quite
artificial.

There are several points that may deserve further investigations.
It would be interesting to study the supersymmetry properties of
these black holes by embedding the lagrangian~\eqref{action} in
some supergravity theory. The extension of the asymptotic symmetry
algebra to a supersymmetric asymptotic symmetry algebra in the
spirit of \cite{Banados:1998pi} would then allow one to fix the
lowest value $\Delta_0$ of $L_0$ undetermined classically and left
ambiguous even after the matching of the entropy with the Cardy
formula. Note that the naive dimensional reduction on a 2-sphere
of the $5d$ minimal supergravity \cite{Gauntlett:2002nw} in which
G\"odel black holes were studied \cite{Gimon:2003ms} does not
admit~\eqref{Sol_gen} as solutions. There are however other
alternatives. Namely, it turns out that the three-dimensional
G\"odel black holes can be promoted to a part of an exact string
theory background \cite{Detournay:2007aa} along the lines of \cite{Israel:2004vv,Detournay:2005fz}, and are in particular
solutions to the low energy effective action for heterotic or type
II superstring theories. It could therefore be instructive to
check if the present asymptotic analysis holds in this latter
theories as well and then study the supersymmetry properties of
these solutions.

\section*{Acknowledgments}

We would like to thank M. Ba\~nados, G. Barnich, S. Carlip, J. Gegenberg, G. Giribet, M. Henneaux, D. Klemm, C. Martinez,  R. Olea, M. Park and Ph. Spindel for enlightening discussions and comments about the topics
discussed in this paper. G.C. thanks the String Theory Group of
the Milano I University, where this work was completed, for its
hospitality. S.D. thanks people at the Centre de Physique
th\'eorique de l'Ecole Polytechnique in Palaiseau for useful
comments, remarks and suggestions. G.C.~is Research Fellow of the National Fund
for Scientific Research, Belgium.  His work is supported in part
by a ``P{\^o}le d'Attraction Interuniversitaire'' (Belgium), by
IISN-Belgium, convention 4.4505.86 and by the European Commission
program MRTN-CT-2004-005104, in which G.C. is associated to
V.U.~Brussel.


\providecommand{\href}[2]{#2}\begingroup\raggedright\endgroup

\end{document}